\documentclass[11pt,twoside]{article}

\usepackage{asp2014}
\pdfoutput=1
\aspSuppressVolSlug
\resetcounters

\begin{document}



\title{Data management and execution systems for the Rubin Observatory Science Pipelines}
\hypersetup{pdftitle={\@title}, pdfauthor={\@author}, pdfkeywords={\@keywords}}

\author{Nate~B.~Lust,$^1$ Tim~Jenness,$^2$ James~F.~Bosch,$^1$ Andrei~Salnikov,$^3$ Nathan~M.~Pease,$^3$ Michelle~Gower,$^4$ Mikolaj~Kowalik,$^4$ Gregory~P.~Dubois-Felsmann,$^5$ Fritz~Mueller,$^3$ and Pim~Schellart$^1$}
\affil{$^1$Department of Astrophysical Sciences, Princeton University, Princeton, NJ 08544, USA}
\affil{$^2$Rubin Observatory Project Office, 950 N.\ Cherry Ave., Tucson, AZ  85719, USA}
\affil{$^3$SLAC National Accelerator Laboratory,  2575 Sand Hill Rd., Menlo Park, CA 94025, USA}
\affil{$^4$NCSA, University of Illinois at Urbana-Champaign, 1205 W.\ Clark St., Urbana, IL 61801, USA}
\affil{$^5$IPAC, California Institute of Technology, MS 100-22, Pasadena, CA 91125, USA}
\paperauthor{Nate~B.~Lust}{}{0000-0002-4122-9384}{Department of Astrophysical Sciences}{}{Princeton}{NJ}{08544}{USA}
\paperauthor{Tim~Jenness}{}{0000-0001-5982-167X}{Rubin Observatory Project Office}{}{Tucson}{AZ}{85719}{USA}
\paperauthor{James~F.~Bosch}{}{0000-0003-2759-5764}{Department of Astrophysical Sciences}{}{Princeton}{NJ}{08544}{USA}
\paperauthor{Andrei~Salnikov}{}{0000-0002-3623-0161}{SLAC National Accelerator Laboratory}{}{Menlo Park}{CA}{94025}{USA}
\paperauthor{Nathan~M.~Pease}{}{0000-0002-9701-5975}{SLAC National Accelerator Laboratory}{}{Menlo Park}{CA}{94025}{USA}
\paperauthor{Michelle~Gower}{}{0000-0001-9513-6987}{NCSA}{}{Urbana}{IL}{61801}{USA}
\paperauthor{Mikolaj~Kowalik}{}{0000-0002-9801-5969}{NCSA}{}{Urbana}{IL}{61801}{USA}
\paperauthor{Gregory~P.~Dubois-Felsmann}{}{0000-0003-1598-6979}{IPAC}{}{Pasadena}{CA}{91125}{USA}
\paperauthor{Fritz~Mueller}{}{0000-0002-7061-4644}{SLAC National Accelerator Laboratory}{}{Menlo Park}{CA}{94025}{USA}
\paperauthor{Pim~Schellart}{}{0000-0002-8324-0880}{Department of Astrophysical Sciences}{}{Princeton}{NJ}{08544}{USA}

\begin{abstract}
    We present the Rubin Observatory system for data storage/retrieval and pipelined code execution. The layer for data storage and retrieval is named the Butler. It consists of a relational database, known as the registry, to keep track of metadata and relations, and a system to manage where the data is located, named the datastore. Together these systems create an abstraction layer that science algorithms can be written against. This abstraction layer manages the complexities of the large data volumes expected and allows algorithms to be written independently, yet be tied together automatically into a coherent processing pipeline. This system consists of tools which execute these pipelines by transforming them into execution graphs which contain concrete data stored in the Butler. The pipeline infrastructure is designed to be scalable in nature, allowing execution on environments ranging from a laptop all the way up to multi-facility data centers. This presentation will focus on the data management aspects as well as an overview on the creation of pipelines and the corresponding execution graphs.
\end{abstract}

\section{Introduction}

\citet{I08_adassxxxii} introduces the core design principles behind the Rubin science processing pipelines \citep{2019ApJ...873..111I}. With this philosophy in mind, we have created systems, known as the Butler and middleware. These systems manage the complexity of storing, retrieving, and organizing large numbers of files. They also store information about the data, such as the relationships between individual units of data. Within the Butler and middleware, an individual unit of data is called a dataset. These systems work as an abstraction that algorithmic code sits on top of such that authors of those algorithms need not concern themselves with the complexity of data management and orchestration with other algorithms.

\section{The Butler and Middleware}
The Butler is a system that knows all about data, so a user or code author does not need to. It knows where the data is stored, the format it is stored in, and the relationships between datasets. Users of this system only need to deal with in-memory Python objects as the Butler manages the appropriate on-disk storage format, using formatters to convert between the two. Furthermore, this abstraction allows us to manage transparent conversions to various in-memory data representations for the same underlying on disk format, i.e.\ PyArrow vs Pandas DataFrames.

The Butler is a high level interface that most users and software make use of. It is made up of two lower-level pieces, the Registry and the Datastore. The Registry is a SQL database compatible with several different implementations. Its job is to maintain all the data about the data. This includes unique identifiers, as well as relationships between the data sets. The Datastore is what actually stores the datasets. It understands the file formats used to store data and their Python in-memory equivalent types. The files themselves may be stored on one or more backends: POSIX, S3, Google cloud services, WebDav, or they may be kept in memory as intermediate objects which are never written to disk.

We refer to the collection of software that interacts with the Butler as the middleware. This software handles everything from organizing user queries to orchestrating the processing of data stored within the Butler. A more in-depth description of the Butler and middleware can be found in \citet{2022SPIE12189E..11J}.

\section{Dimensions}
As mentioned Dimensions are a way to organize, localize, and identify data within a large volume of possibilities. The set of dimensions is fixed for a given project before any data is tracked. This can be thought of as the ``data model'' of the project, a complete description of all the identifiers that might be associated with any given dataset.

Dimensions additionally can relate to each other. For instance the Rubin dimension system includes dimensions of \texttt{exposure} and \texttt{detector}. Because Rubin's \texttt{instrument}, it self a dimension,  has 189 individual \texttt{detectors}, a given \texttt{exposure} implies 189 corresponding points along the \texttt{detector} dimension. Alternatively, it is possible to investigate for a given \texttt{detector}, how does data vary along the \texttt{exposure} dimension (and thus is a proxy for time). Dimensions may relate to each other in logical ways, such as the previous example, as well as spatially, or temporally.

\section{Datasets}
In order to do its job of managing datasets, the Butler's view of what a dataset is is more than just a bunch of bits on a disk somewhere. Each dataset tracked by the Butler is a composition of three pieces, a dataset type, a data ID, and a collection.

The dataset type describes the ``what this is'' of a dataset. It has a name to use as an identifier, a storage class that maps on-disk formats into in-memory Python objects, and a subset of Dimensions. This subset is used as identifier keys that describe how to address the given dataset type, as outlined above.

A data ID is akin to the address for a given dataset type. It is a mapping of the dataset type's Dimensions to corresponding concrete values. These form a ``data coordinate'' within the set of all dimensions, just like an x,y,z does within a Cartesian space.

Collections are a way to group data together. A collection is simply a name, but it allows two or more datasets that may exist at the same ``data coordinate'' to both exist within the Butler without any collision or ambiguity. Within a given collection, however, there can only be one dataset at a given coordinate.

\section{Making use of Data in the Butler}
All algorithms in the Rubin processing pipeline, are special objects called Tasks. Tasks are a single logical job, composable inside other tasks, and have a schema for configuration which is separately defined and able to be persisted. All tasks operate on in-memory data products, normally supplied from another task.

There is a special subclass of tasks, called PipelineTasks, that get their in-memory data by supplying information to the middleware so that they may do I/O with the butler on the task's behalf. PipelineTasks declare an interface of ``connections'', which are the task's desired input / output dataset types, along with the connections' dimensions. They also declare a set of dimensions the task will operate on. The combination of these dimensions with the connection dimensions enables the middleware to decide how to supply inputs to the task. For example, extending the example above, if the task declares its dimensions to be \texttt{exposure}, and an input to be \texttt{exposure}, \texttt{detector} then a single invocation of the task will be supplied 189 inputs (one for each \texttt{detector}). In contrast if the task's dimensions are \texttt{exposure}, \texttt{detector} as well, then there will be 189 separate invocations of the task, each receiving one input.

\section{Pipelines and Graphs}
Pipeline files, written in YAML syntax, provide a way of grouping multiple PipelineTasks together. In a Pipeline, each PipelineTask is declared with a unique label, and may contain configuration overrides to a task's default values. Pipelines are designed for a purpose, such as preparing calibrations, or processing data from a specific instrument. As such, it is possible that a PipelineTask may appear in more than one Pipeline with different configurations. Using unique labels, it is also possible that the same PipelineTask appears multiple times with unique configurations within one Pipeline.

The middleware turns these pipelines into directed acyclic graphs (DAG) in execution dependency order. It does this by examining the connections declared by each task in the Pipeline, in the context of its configuration, which may modify said connections. Because the ordering is done by the middleware, the order in which PipelineTasks are declared within a pipeline is not important. This system insulates task authors from the overall complexity, and provides maximum reusability, as the system is able to use a task in widely varying contexts. It also allows authors to work in a system they are more familiar with from software development where they need only be concerned with interfaces rather than the whole application.

Once the middleware has constructed a pipeline graph, that graph can be used as an input in further processing steps. The pipeline graph is taken along with a Butler location, user specified collections of data, and user specified data constraints.  The middleware then examines the input nodes in a pipeline graph to determine the set of data in the Butler that matches the node's Dataset type and Dimensions. Using these inputs, and information the Butler knows about the relationships between dimensions, the middleware is able fill in all the remaining nodes based on what datasets it predicts will be created, consumed, and output. Each of these new nodes is called a Quantum, and represents an individual invocation of a task. The collection of all the nodes in the new DAG is known as the QuantumGraph. Creation of this graph is done before any data processing takes place, allowing for appropriate resources to be allocated or scheduled. This same code can be used to process data on a wide range of computing platforms, from a laptop with hundreds of Quanta, up to production runs on compute clusters on physical hardware or in the cloud as outlined in \citet{P52_adassxxxii}.

\section{Conclusions}
The Butler and middleware allow Rubin to abstract all data access from task authors or end users. Abstracting data access also allows allows data processing to be abstracted using PipelineTasks and Pipelines. These abstractions allow more rapid, orthogonal development to take place, as well as providing a mechanism to use the same code base to scale from individual development needs all the way up to running the complete survey. These systems are being used in processing Subaru's Hyper Suprime-Cam PDR4, replacing the older Rubin middleware used for previous releases, e.g. \citet{2018PASJ...70S...5B}. It is also being used in NASA's SPHEREx mission, which has its own data model. The code is all open source and available from the Python package index.

\acknowledgments This material or work is supported in part by the National Science Foundation through Cooperative Agreement AST-1258333 and Cooperative Support Agreement AST1836783 managed by the Association of Universities for Research in Astronomy (AURA), and the Department of Energy under Contract No.\ DE-AC02-76SF00515 with the SLAC National Accelerator Laboratory managed by Stanford University.


\end{document}